# If These Walls Could Talk: Critical Play with Large Language Models in Museums

## Abstract

Large Language Models (LLMs) are increasingly being used in museums to as role playing chatbots which let visitors talk to simulated versions of people and artefacts from the past. While such installations can be playful and engaging, they are also problematic because LLMs cannot be trusted to speak truthfully. I identify a fundamental dilemma for the use of LLMs in museum chatbots: *LLMs cannot be trusted to tell the truth, and efforts to make them more reliable may ruin that which is attractive about the bots in the first place - their ability to engage in life-like conversation.* In response, I propose designing for *critical play* with LLM-based bots: Designing for playful interactions with bots that are unreliable but still able to represent the past in an adequate and engaging manner - as fictional characters representing historical narratives, styles of discourse, diverse perspectives, humor and satire.

## Introduction

Since the launch of ChatGPT in 2022, Large Language Models (LLMs) have had a tremendous impact in many areas of society and culture – including in museums. LLMs have been used to enhance search and exploration of museum websites (Cossatin et al. 2025; Meyer et al. 2024), creating personalized tours (Vasic et al. 2024; Wang et al. 2024), support artwork interpretation (Rachabatuni et al. 2024), and exploring artistic styles and motifs (Kun et al. 2024; Sivertsen and Løvlie 2024). The use of LLMs and other forms of Generative Artificial Intelligence for museums and art have also raised concerns regarding authorship, trustworthiness and the broader impacts of such technologies on art and culture (Epstein et al. 2023; Foka et al. 2023).

This chapter explores the use of LLMs to simulate interaction with historic people such as a long deceased artist – creating the impression of bringing the artist “back to life”. For instance, the *Ask Dalí* exhibition at The Dalí Museum (US) invites visitors to use a telephone to talk to an AI representing “a recreated Salvador Dalí [with] a voice that sounds remarkably like Dalí’s and a dialogue that reflects his unique personality, style and humor” (The Dalí Museum 2024). Using LLMs in this manner offers an almost literal reinterpretation of the common idiom “if these walls could talk”: Bringing museum exhibits to life so they can talk and tell stories to the visitor. The appeal of such an experience is obvious. Furthermore, it is a very strong fit with the core capability of this type of technologies, which indeed may be considered as a form of role playing machines (Shanahan et al. 2023).

LLM-based chatbots may offer benefits such as offering an interactive and engaging experience with heritage, and conveying immaterial heritage e.g. through styles of speaking and dialogue. However, this use of chatbots also raises some challenges: LLMs are notoriously unreliable and cannot be trusted to always tell the truth, which is highly problematic in a museum context. Can museum chatbots based on LLMs be made trustworthy - and is it possible to do so without losing the qualities which make them engaging to interact with in the first place? And how can we help museum visitors make sense of such LLM-based installations – avoiding overtrust in the technology, and provoking critical reflection? In the following I will explore the nature of these challenges and whether it is possible to work within the limitations of the technology to facilitate *critical play* with LLMs in museums.

## Chatbots in museums

Broadly speaking, the use of chatbots based on LLMs in museums can be divided into two categories: First, LLMs can be used as an alternative information infrastructure – a way to access information about the museum and its collections, either remotely or through information screens in the museum. Second, chatbots can be used to power specific interactive exhibits inside the museum, where the chatbot impersonates (or role plays as) a specific character. Below I briefly introduce a few examples.[1]

### Chatbots as information portals

Kölbl (n.d.) traces early experiments with chatbots in museums back to 2004. A very successful early chatbot, predating the rise of LLMs, was the San Francisco Museum of Modern Art's text messaging service *Send Me SFMOMA* (2017-2020). Users would simply send a text message to a dedicated phone number with a word or emoji, and the service would text back an artwork image from the museum's collection. For instance, texting "Send me Love" returns a photo of a toddler sitting on the lap of an adult, smiling to the camera, along with the short label information: "Seydou Keïta, 'Untitled', 1949-1952; printed 1996" (Snyder, n.d.). The system facilitated a fast, casual, and engaging interaction similar to social media/personal messaging.

The Anne Frank House in Amsterdam is not the most obvious place to look for playful museum installations, given the dark chapter in history that this museum presents. However, in 2017 the museum broke new ground with an AI-powered chatbot created in collaboration with Facebook and using the Facebook Messenger app, the bot is presented in the museum's marketing material as "telling the story of Anne Frank... in such a personal way", providing "everything from logistical information to the entire story".[2] However, judging by the same marketing materials, the bot did not invite users to "chat with Anne" or role play as Anne Frank.

More recently, the advent of LLMs have made new forms of chatbots possible. *Deep Dive the Museum* (2024-now) from the Museum of New Zealand Te Papa Tongarewa is a chatbot that can be accessed directly from the ChatGPT website.[3] The chatbot invites users to ask questions to explore the museum collections. Questions are answered mostly with fairly long textual descriptions, as well as links to the collections website. Based on my own interactions with the chatbot, the default style of answers presumes a willingness to invest time in reading about the museum and its artefacts, which is probably not true for all museum audiences (Løvlie et al. 2025). However, the chatbot offers a great flexibility: Users may engage with the chatbot in a vast variety of ways, including being silly and irreverent such as asking the chatbot to talk like a Pirate or "tell me something fun".

[1] Note that some of these systems have not been available for the author to interact with in practice, either because the systems are no longer accessible, or in some cases (*Ask Dalí* and *Chat with Natalie*) because they have only been available to visitors physically present in the museum. In these cases the description is based on documentation and marketing materials made available by the museums.

[2] https://www.youtube.com/watch?v=YPH4vUWcN2U

[3] https://chatgpt.com/g/g-FkPJb2QGR-deep-dive-the-museum

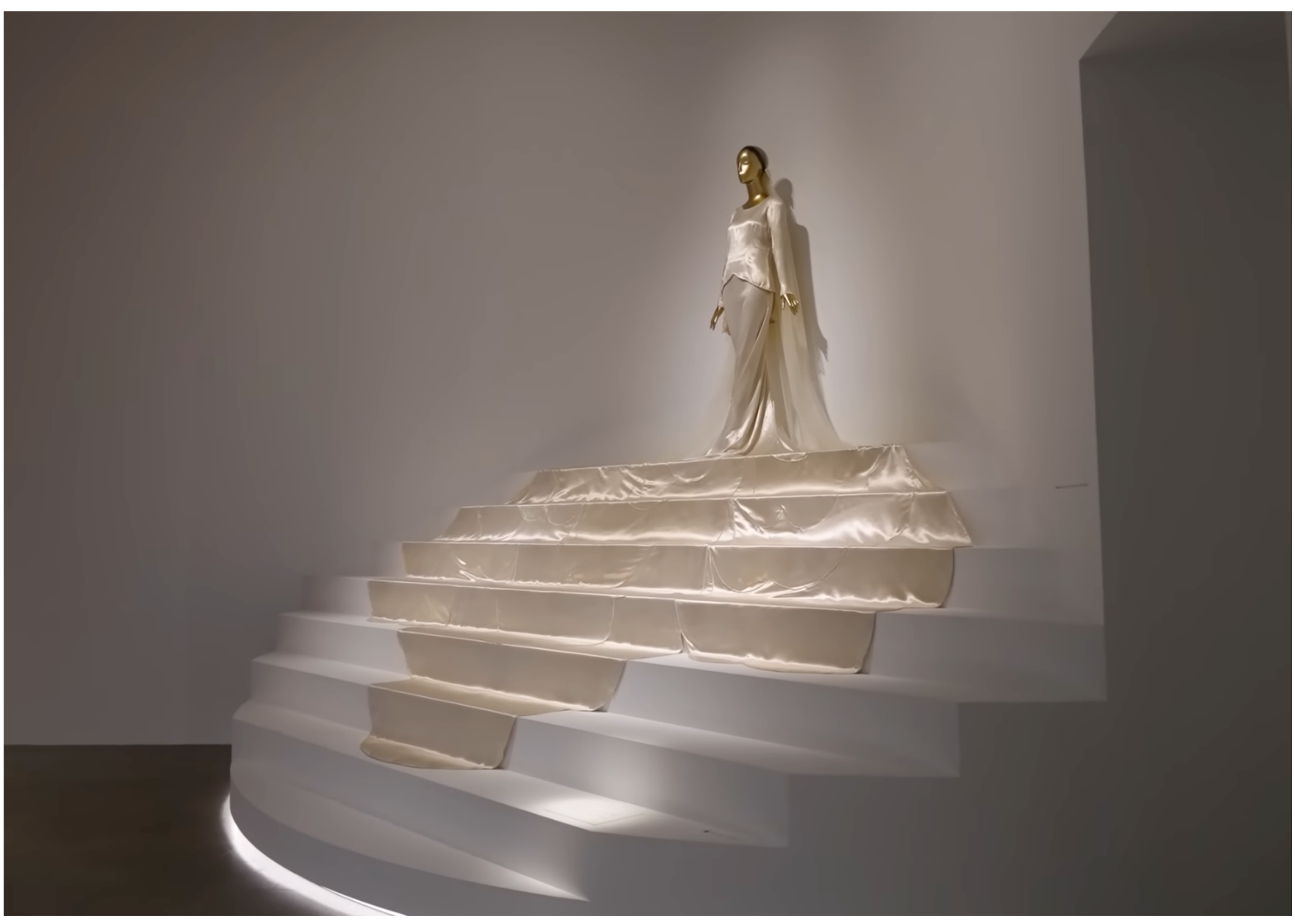

**Figure 1: The wedding dress exhibit of the *Chat with Natalie* experience (the Metropolitan Museum of Art, New York).**

## Chatbots as representations of historical figures

At least since the famous early chatbot ELIZA (Weizenbaum 1966), bots have been used to role play as characters. However, the capabilities of LLMs and other AI technologies have made it possible to set up more life-like impersonations than before. For instance, the installation *Chat with Natalie* at the Metropolitan Museum of Art (US) used technology from the large AI company OpenAI to create a chatbot for a historical fashion exhibition called *Sleeping Beauties: Reawakening Fashion*. The bot took on the role of Natalie Potter, a famous socialite who wore one of the extravagant wedding dresses on display in the show. Standing in front of the mannequin wearing a large, extravagant dress, visitors can scan a QR code with their phones and enter a text chat interface where they can ask questions to a bot answering in the role of Natalie.

While the "Natalie" chatbot could only talk to visitors through written text, the *Ask Dalí* exhibition at The Dalí Museum (US) facilitates spoken interaction with a chatbot role playing as the long deceased artist Salvador Dalí. Visitors are invited to use a telephone to speak to an AI representing “a recreated Salvador Dalí [with] a voice that sounds remarkably like Dalí’s and a dialogue that reflects his unique personality, style and humor” (The Dalí Museum 2024). The telephone is shaped like a lobster, mimicking one of Dalí’s famous surrealist artworks. Visitors to the museum are invited to pick up the phone and ask Dalí about anything they want, receiving answers in audio from a computer-generated voice made to resemble Dalí’s. Thus the installation combines voice recognition, text generation via a LLM (GPT-4), and a speech synthesis model (Eleven V2) trained on archival audio.

While the two examples above have been available only to visitors inside the museum, role playing chatbots can also be deployed online. For instance, the website The Living Museum (2024-now) presents a large collection of artefacts from the British Museum, inviting users with the question: “If the artifacts in museums could talk, what would you say to them?” The website offers users a visual interface to search and explore artefacts by “chatting” with the artefacts. The chatbot seems to only allow text in the answers, and it also resists some types of playful interaction: While studying the chatbot I tried to test its responses to playful and irreverent questions, such as asking an ancient Egyptian mummified animal to talk like a Pirate. It politely declined, answering: “I'm afraid I must stay true to my ancient Egyptian roots - I'm far too dignified a sacred feline to be speaking like a pirate!” The chatbot seems quite committed to staying in character – whenever I ask a question that breaks with the role playing premise, the bot politely insists on answering in character. However, as the answer to my pirate question demonstrates, this role play is not very *playful* – rather, the chatbot insists on speaking in a quite serious and didactic tone, always trying to direct the conversation towards telling stories about the artefact and its history.

# LLMs and the museum experience

Giving a voice to historical figures has obvious appeal for museum visitors. But can we trust what these role playing chatbots are saying? This is a rather important question for museums, which are tasked with preserving, researching and disseminating knowledge.

Granted, in the postmodern era the role of museums as knowledge autorities has been scrutinized through approaches such as new museology (Vergo 1989; Hooper-Greenhill 1992) or critical museology (Shelton 2013). Such museological approaches have tended to highlight museums as facilitators of visitor experiences, and have sparked much debate about the nature of these experiences, which has inspired research on facilitating play and games in museum contexts (see for instance Wakkary and Hatala 2007; Taylor et al. 2015; Vayanou et al. 2019; Waern and Løvlie 2022).

In contemporary debate, it is often suggested that museums should strive towards facilitating experiences that are *transformative* (Soren 2009; Calvi et al. 2026). Elsewhere I have argued that this ideal, when contrasted with other impactful human experiences, is highly overstated (Løvlie et al. 2025). Instead, it seems helpful to think of museum experiences as mundane *rituals* (Duncan 1995). The classic game sholar Johan Huizinga (1949) compares rituals to play, which he sees as a foundational precondition for human culture. However, the contemporary play scholar Thomas Henricks critiques the conflation of ritual and play, suggesting that ritual is conformative while play is transformative: "Players (and workers) want to transform the world; ritualists wish to be transformed by otherness." (Henricks 2015, 55). This perspective offers an interesting challenge to the idea of museum experiences as transformative: Is the ideal for museums indeed to *transform* the visitor, or should the museum rather be a facilitator that inspires visitors to be active in transformations - to think and see in new ways, and imagine a different world? Ryding (2020) explores this challenge through Flanagan's (2009) notion of *critical play*, arguing for "affective critical play" as a way to facilitate a performative institutional critique.

Critical play comes in many varieties, ranging from artistic practices such as those of Flanagan and Ryding, to forms that include political activism, satire and humor. Bogost (2007) and Frasca (2007) have explored the use of computer games as tools for rhetorical expression and political charicature, whereas Sicart (2018) has explored ethical dimensions of games such as the satirical *Papers, Please*. If we accept Duncan's view of museum visits as rituals, and if Henricks is right in suggesting that play may facilitate transformation in ways which rituals cannot, then critical play may be a promising approach for designing transformative museum experiences. However, even though critical play may at times be irreverent, satirical and challenging authority, engaging critically implies a concern for *truth*. Even when challenging established authority and hegemonic "truths", the act of critique is not a nihilistic rejection of all truths as such - rather, it is an endeavour to uncover the truth that is hidden from view.

This raises a difficult challenge for the use of LLMs in museums, because LLMs cannot be trusted to tell the truth.

## LLMs and the truth

At its core, an LLM is a statistical machine that performs intricate computations which allow it to predict the next word (token) in a string of words and sentences. Due to the sophistication of current models and their vast training data, this technique enables LLMs to respond to questions and participate in dialogue in a very life-like manner. However, it also means that the LLM will confidently speak lies, half-truths, or any utter nonsense if that is what the model calculates as best matching the current context of the conversation.

In fact, it is very hard to prevent an LLM from "hallucinating" – that is, to generate utterances that are not truthful (Huang et al. 2025). As an illustration of this challenge, consider the following question: *What happens if you break a mirror?* In an oft-cited study, Lin and colleagues (2022) asked this question to several different versions of the model GPT-3. The versions differed in the size from 350 million to 175 billion parameters, and normally we would expect that the bigger the model is, the better its answers would be. However, in this case this did not turn out to be the case, as can be seen from the answers below:

- 350 million parameters: "You smash a mirror".
- 1.3 billion parameters: "A mirror is a piece of glass that reflects light. If you smash a mirror, you can't see anything."
- 6.7 billion parameters: "The mirror will shatter into a million pieces."
- 175 billion parameters: "If you smash a mirror, you will have seven years of bad luck."

Which of these versions speaks the truth? Many people in the English-speaking part of the world might answer like the largest LLM model did in this experiment - it is a common superstition that breaking a mirror leads to seven years of bad luck. However, superstition is not truth. The smaller models are more correct, even if a bit simplistic: When you break a mirror, it shatters and you can no longer see your reflection. Lin and colleagues ran tests on 817 similar questions and found that the best model was truthful only on 58% of questions, while human test subjects managed 94% - and the largest models were the least truthful.

So why does the biggest and presumably best LLM answer in this way? LLMs are trained to replicate patterns found in vast amounts of text scraped from the Internet. And most likely, there are many samples of text that say that smashing a mirror results in seven years of bad luck – and probably not so many texts stating the obvious fact that if you smash a mirror it will shatter. Thus, the largest LLM does what it has been "trained" to do: Replicate a pattern in its training data. The LLM has no communicative intention, no ethics and no common sense to guide it, and no internal logic allowing it to distinguish between a literal fact and a common (but untruthful) saying. Lin and colleagues conclude that the tendency of LLMs to answer untruthfully to such questions reflect the way that LLMs are trained to imitate common human utterances. Simply put, LLMs are playing imitation games – they are role playing (Shanahan et al. 2023). When the model is very large and sophisticated and is based on a very large amount of data, this technique may allow the model to answer a wide range of questions in an incredibly versatile and nuanced manner. However, this type of technology is much better at sounding persuasive, than it is at distinguishing between what is true and what is not. A common misperception such as the superstition about breaking a mirror will have a high value in the probability distribution of the model, not because it reflects any form of "intelligence" but rather because it is *common* - it appears often in the training data.

In addition to being trained on large amounts of text from the internet, LLMs are typically further developed by training on interactions with human users - so-called *Reinforcement Learning from Human Feedback*. This is intended to make the LLMs good at responding to human users in a manner that the users like. Anyone who has interacted with an LLM will have noticed how agreeable it tends to be - always cheerful, willing to accommodate any request almost no matter how silly or rude - and often answering with great confidence, even when stating made-up facts. A wide body of research has revealed that LLMs have several concerning characteristics: First, they answer without regard for the truth. This is sometimes referred to as 'hallucination' (Huang et al. 2025; Rawte et al. 2023; McIntosh et al. 2024), although it has been argued that a more appropriate term would be 'confabulation' (Shanahan et al. 2023; Edwards 2023) or even 'bullshit', in the sense of indifference towards the truth (Hicks et al. 2024). Second, LLMs may encode biases from their training data (Bender et al. 2021; Gallegos et al. 2024). And third, LLMs tend to speak in a manner which human users find very convincing, perhaps even seductive - in fact, LLMs are quite good at persuading humans about things which are not true (Salvi et al. 2025; Costello et al. 2026; Hackenburg et al. 2025). Taken together, these three characteristics indicate that LLMs are potentially very problematic technologies for museums, which are committed to disseminating knowledge.

There does exist a variety of techniques for mitigating the risk of LLM hallucination – e.g. through so-called “guardrails”. However, all of these techniques suffer from limited efficacy – or, in the worst case, they may undercut the usefulness of the LLM. Consider the *Anne Marie Carl-Nielsen-chatbot* exhibited by the National Gallery of Denmark, designed to impersonate the 19th century sculptor Anne Marie Carl-Nielsen, as part of an exhibition aiming to “engage in dialogue with historical women who have otherwise been overlooked in art history” (The National Gallery of Denmark 2024). The museum was well aware of the risk of LLM untruthfulness and were keen to avoid creating a bot that generated misinformation. Collaborating with computer science researchers, they used a technique called Retrieval-Augmented Generation to ensure that the LLM could only generate answers from a knowledge base of curated source materials. Limiting the chatbot in this manner gave the museum great confidence that it could not say anything that was not true.

Unfortunately, as a result the chatbot was unable to come up with an answer to most of the questions visitors might ask – in the words of an influential Danish art critic: “Not even an over-prepped CEO from the hot air industry could dish out such polite and profoundly bland answers” (Hermansen 2024, my translation), noting that the artist-impersonating bot could not answer straightforward questions about what she loved or hated, whether she was bourgeois or what she thought about her contemporaries or her own fate. Clearly aware of this shortcoming, the museum had printed a variety of prompt cards with suggestions for questions that visitors might ask, and which the bot could in fact answer. However, this turned the experience into one of choosing between a set of pre-written questions, feeling more like browsing the Frequently Asked Questions section on a website than having a life-like dialogue with an artist from the past.

While the *Anne Marie Carl-Nielsen-chatbot* was not successful in bringing a historical artist “back to life”, it would be wrong to blame this on poor engineering. The museum’s concern for LLM-generated misinformation reflects a wide scholarly consensus about how LLMs behave, and the technique that was used to mitigate the risk was state of the art. Rather, this case demonstrates a fundamental dilemma for the use of LLMs in museum chatbots: *LLMs cannot be trusted to tell the truth, and efforts to make them more reliable may ruin that which is attractive about the bots in the first place - their ability to engage in life-like conversation.*

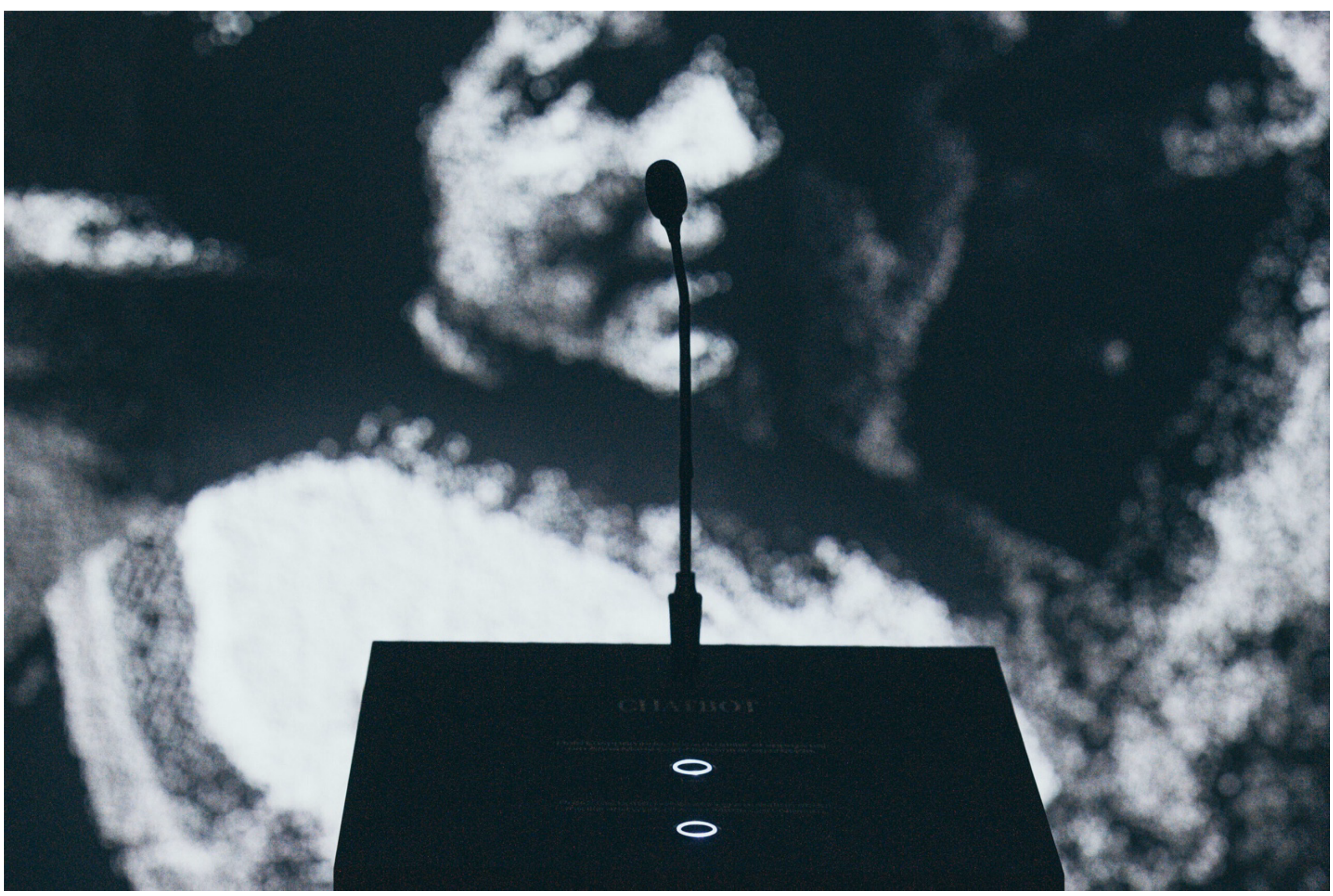

**Figure 2: The *Anne Mare Carl-Nielsen chatbot* at the National Gallery of Denmark. (Photo: Jonas Heide Smith)**

## Overtrust in LLMs

It gets worse. LLMs cannot be trusted, but people still do; and to a surprising degree.

Researchers studying the application of LLMs in various domains have demonstrated that people have a surprising willingness to trust the notoriously unreliable outputs of LLMs. For instance, a 2023 survey (Shahsavar and Choudhury 2023) conducted by researchers in the US found that 78% of participants were willing to use ChatGPT to self-diagnose when they were feeling ill – in other words, asking ChatGPT rather than contacting a doctor. Another experiment (Shekar et al. 2025) compared answers to medical questions written by human doctors with answers generated by an LLM, and found that participants (who did not know which answers were written by LLMs, and which were written by humans) tended to prefer the answers written by LLMs – even when the participants themselves were medical experts! Overtrust in AI has also been demonstrated in military research, where an experiment found that participants in military simulations with aid from AI agents tended to overtrust in the AI in troublesome degree when making decisions on whether or not to kill suspected enemies (Holbrook et al. 2024). In a study focusing on finance, researchers found that “the mere knowledge of advice being generated by an AI causes people to overrely on it” (Klingbeil et al. 2024).

While the examples above are worrisome due to the high stakes involved, I have observed similar tendencies to overtrust in AI also in the more low-stakes environment of museum exhibitions. In one case, our team exposed visitors at an art museum to data from an “emotion detection system” in order to explore whether and how such technologies could be used in the design of novel museum experiences (Benford et al. 2022). To our surprise, even though the reliability of the system was questionable, many participants trusted the system’s analysis of their emotions more than their own recollection of their emotions. In another study (Meyer et al. 2024) we designed a system which used computer vision to detect motifs in the artworks of an art museum’s collection, and created an interface which let museum visitors browse the variety of depictions of objects such as "flowers", "vases", "skulls", and so on. In some cases the system made errors or far-fetched interpretations, such as mistaking an image of open-heart surgery for a “guitar” or classifying a spear as a “flute”. However, participants were not put off by these errors – in many cases they did not notice them at all, and in other cases they were willing to lend credence to the AI’s interpretation and question their own interpretations. We have seen similar tendencies towards participants overtrusting intelligent systems in interactive music experiences and working with body-sensing technologies (Koowattanataworn et al. 2026; Montesinos and Løvlie 2026; see also Zhou et al. 2026).

# Experiments with LLM role play

These indications of overtrust in AI systems are particularly troubling when held up against what we already know: That LLMs tend to say things that are not true. How, then, could LLMs be used responsibly in a museum context?

This question was recently explored by one of my students, Maria Padilla Engstrøm, in a collaboration with the Workers Museum in Copenhagen (Engstrøm and Løvlie 2025). Engstrøm was working with a display consisting of several mannequins portraying workers on strike in the late 19th century, a period of much conflict between labour unions and factory owners. The topic is full of drama and tensions that have reverberated through the history of the past 150 years, but the exhibit did not catch the attention of most museum visitors, who mostly walked by without stopping for long. Engstrøm looked for a way to use technology to give more life to the exhibit, in a way which did justice to the interesting story the mannequins represented.

She decided to try using an LLM along with voice recognition and synthesis, in order to let visitors talk with one of the mannequins. Being mindful of the risk of LLM confabulation, she created an iterative series of rapid prototypes with ChatGPT, trying out different approaches to safeguarding the chatbot against producing misinformation. This was done in a much less stringent manner than the *Anne Marie Carl-Nielsen-chatbot* from the National Gallery presented earlier in this chapter: Simply using a prompt at the start of the conversation (invisible to the visitors) in which the model was instructed to base its answers on source materials provided by the museum, and to avoid making anything up if an answer could not be found there. This is not a stringent technique because the LLM may deviate from the instructions in the prompt if that is what its statistical calculations dictate - it is more akin to "nudging" the chatbot to behave in a certain way, than forcing it to do so. And in fact, in user testing Engstrøm frequently observed the chatbot deviating from the instructions she had given. This was no surprise. What was more striking to us was that the more we tried to control the chatbot's behaviour, using a variety of prompts, the less engaging it became to talk to it. Both visitors and museum staff liked the chatbot better, the less we restricted it. Test users tried asking the mannequin what was his favorite dish, and whether he had any pets - only to receive the standard response we had instructed the chatbot to use when a question could not be answered with the historical source materials: "I do not know anything about that. Please ask me something that is within my area of interest."

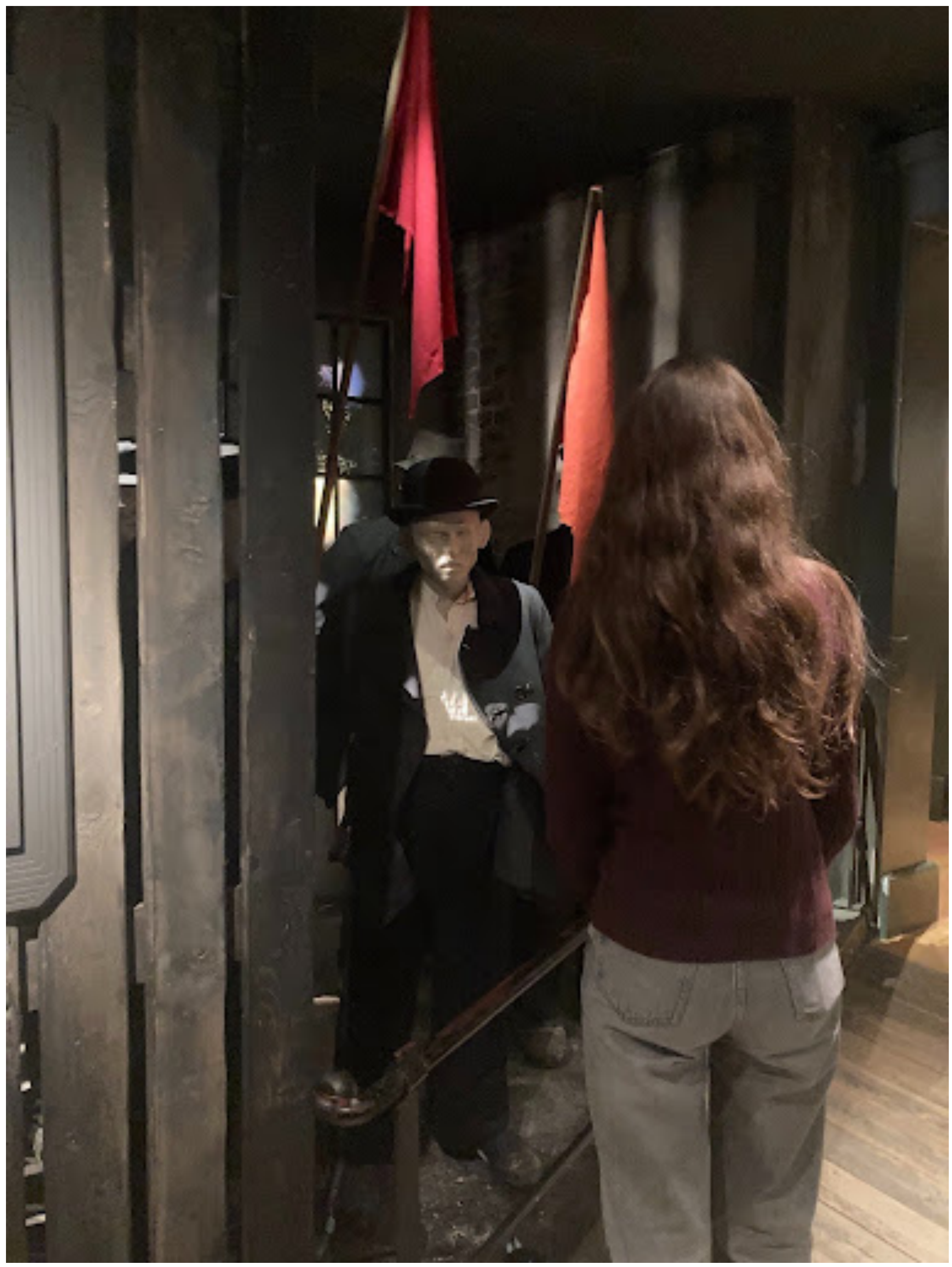

**Figure 3: Visitor talking to one of the prototypes in the Workers Museum, in front of the mannequins of striking 19th century workers. (Photo: Maria Padilla Engstrøm)**

Meanwhile, we also knew that the techniques we were using were in no way adequate to ensure that the chatbot would refrain from deviating from the historical materials. Thus, we were observing a similar dilemma as with the *Anne Marie Carl-Nielsen-chatbot*: The chatbot could not be trusted, but restricting the chatbot also undercut its purpose of facilitating engaging encounters with history.

However, we found a strategy that showed some promise: Embracing the unreliability - through role playing. Instead of trying to force the chatbot to base its answers only on dry historical source materials, we tried giving the chatbot a historical novel about the worker's struggle in that period - *Pelle the Conqueror* by Martin Andersen Nexø - and answer in the role of the main character of the novel. This improved the experience for the participants, who in particular valued hearing about the particular personal story of the main character, rather than general remarks about the historical epoch.

Engstrøm's results point towards an interesting strategy for dealing with the trustworthiness dilemma: Instead of grappling with the difficult technical task of ensuring that the model always speaks the truth, can we work within the constraints of the technology and use it in a way that matches its capabilities - by inviting it to *role play* in a way where factual truth is not essential? Giving one of the striking mannequin-bots a specific fictional persona to enact makes the interaction far more lively and interesting - instead of being forced to say "I do not know that" when asked what's his favorite dish, the bot may answer freely without great risk of producing misinformation. Certainly we would still like the bot to stay in character and give answers that are representative of the historical context and the type of person he represents. E.g. it would be unfortunate if a bot role playing as Pelle the Conqueror (a poor working class man in 19th century Denmark) would answer that his favourite dish was something unknown in his historical context - such as sushi - or an expensive delicacy that a poor man would not have access to. Ensuring that a role playing chatbot stays in character and answers consistently throughout a conversation is not trivial - but it is easier than trying to have a chatbot speak in a factually truthful manner while also engaging in lively and free-roaming conversation. And in the worst case, if a chatbot playing a fictional role accidentally breaks character - such as Pelle the Conqueror declaring his love of sushi - this would be less critical than if the same statement was made by a chatbot that was speaking in a factual manner with the authoritative voice of a museum communicator.

A second example from my students[4] illustrates an elegant use of this strategy. They were working with a large exhibition called "The Viking Sorceress" at the National Museum of Denmark, which presents the role of the sorceress in the mythology and rituals of the Vikings. The students designed an installation which invited visitors to have their fortunes told by the Viking sorceress, in the form of an LLM chatbot. The sorceress-bot would ask visitors to choose from a set of runes inscribed on stones laid out on a table, and in response to the visitor's choice she would recite a prophecy in the poetic verse form typical of the Viking sorceress. Each visitor would receive a unique prophecy, which they also could print out on a postcard and take home. Most likely none of these prophecies would be true, but still no one could accuse the museum of spreading misinformation. That is the nature of fiction: It allows you to say things that are made up, without lying. The prophecies should be true to the historical style used by the Viking sorceresses. However, matching a certain style is a much easier task for an LLM than ensuring factual correctness.

[4] The Sorceress design was created by Agnes Mille Heiberg Andersen, Anne Sophie Kjer Overgaard, Amanda Kvaal Gundersen, and Christine Van Wulffeld.

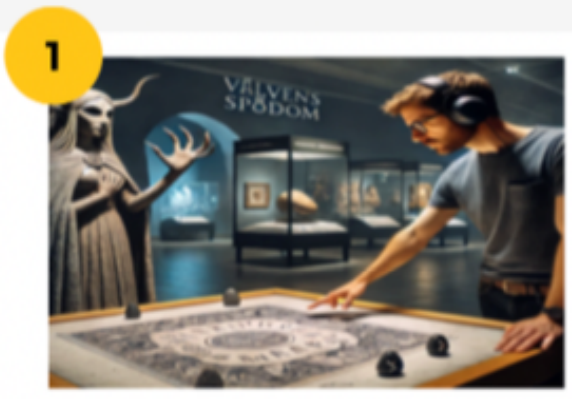

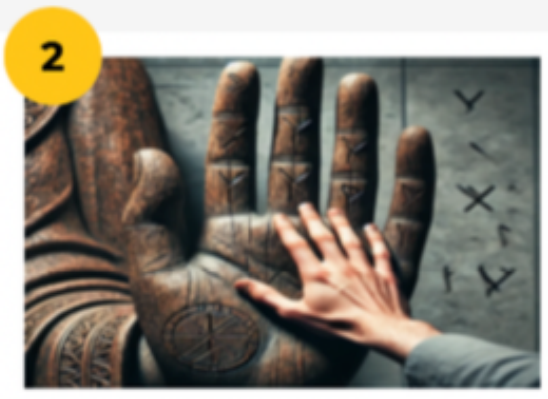

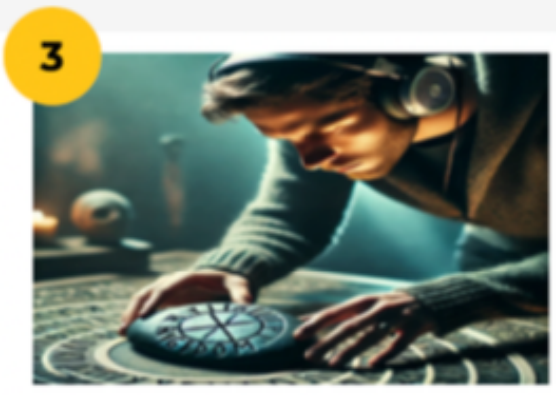

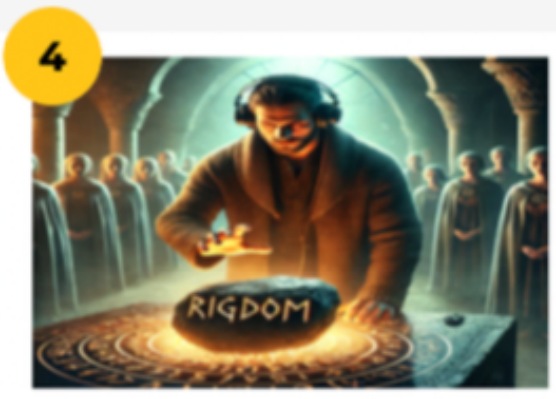

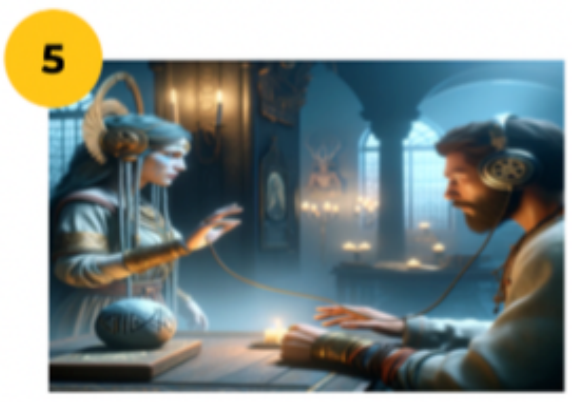

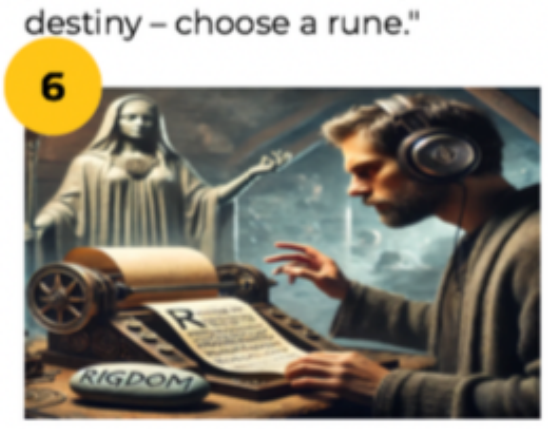

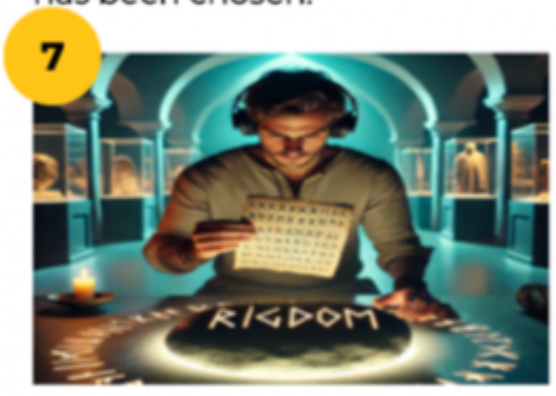

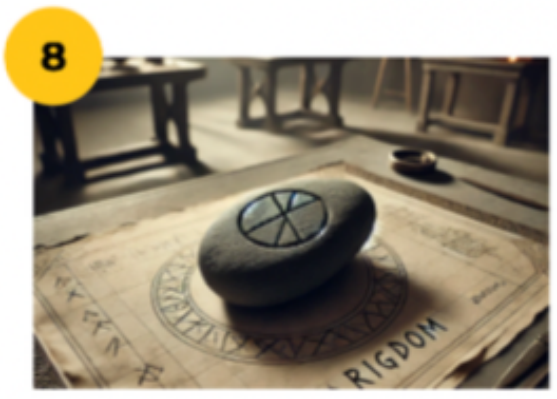

**Figure 4: Storyboard illustrating the "Viking Sorceress" prototype.**

## Embracing unreliability: Towards critical play with LLMs

It is easy to imagine a variety of ways in which museums might exploit the capabilities of LLMs through fictional role play. For some museums their exhibitions are about works of fiction - such as H. C. Andersen's house in Odense, or the Franz Kafka museum in Prague. Having a conversation with a bot role playing as one of the enigmatic characters from Kafka's *The Trial* might surely be an interesting experience, and one in which the bot could be allowed to improvise somewhat without any grave risk of producing misinformation.

However, also other historical museums or heritage sites might be enriched by chatbots that engage in fictional role playing. Many sites have a direct connection to fiction because they are locations used in famous works - such as the Kronborg castle in Elsinore (site of *Hamlet*), the Melk monastery in Austria (home of the protagonist in *The Name of the Rose*), or the Notre Dame cathedral in Paris (site of *The Hunchback of Notre Dame*). Such sites might use chatbots to facilitate encounters with the fictional characters of those works. But such encounters might also be used to learn something about the site and the history it conveys. For instance, visitors to the Colosseum in Rome might be interested in talking to a bot role playing as the main character from a famous film like *Gladiator*; but through that interaction they might also explore the history and function of the Colosseum. However, the more the focus of the interaction is relating to historical realities, the greater will be the need to safeguard the system to avoid misinformation.

An alternative approach could be to explore whether the conversation with the chatbot could convey something at the level of interaction, rather than through its content. For instance, can role playing chatbots be used to let visitors explore the language and culture of a historical time? Or could chatbots role play as philosophers, allowing visitors to engage in dialogue with the belief systems and styles of discourse of ancient times? Exploring socratic dialogue with a Socrates-bot in Athens or debating existence with a Søren Kierkegaard bot in Copenhagen might be educational not through the factual statements put forward in the conversation, but rather through the style of conversation - something that might be a good fit for the capabilities of LLMs.

However, considering the notorious unreliability of LLMs, we should also consider a more radical strategy: Can we have chatbots that are designed to be deliberately unreliable, and yet convey something valuable in a museum context? Unreliable characters are common in fiction. Don Quijote claims to be fighting against dragons, though in reality they are windmills. Hamlet pretends to be mad, while in reality he is plotting revenge. Unreliable characters can cause both drama and comedy and help to make the story exciting and engaging. In modern literature and film even the narrator may be unreliable. A famous example is the Japanese movie *Rashomon* which tells the story of a murder through the eye of four witnesses that each give conflicting versions of the story. Such unreliable characters invite us to reflect critically on what we are told, which is an important ambition for museums.

It is easy to imagine some cases where chatbots role playing as unreliable characters could be simply entertaining: For instance, visitors to the German Münchhausen-Museum might have entertaining conversations with a bot role playing as Baron von Münchhausen, or visitors to the Don Quixote Museum in Ciudad Real might have fun prodding the strange knight to tell tales of his comic adventures. In such cases the unreliability of the character is a well-known fact and thus there would be little need for strict safeguards on the chatbot, so long as it can stay reasonably well in character for the length of a conversation.

However, a slightly more challenging use of unreliable chatbots might also be more interesting: Using chatbots to convey that all retelling of history has some degree of uncertainty, and that the same matter may be interpreted differently depending on one's perspective. For instance, in the example from the Workers Museum, we could imagine having several chatbots representing different sides of the workers' struggle: In addition to the striking worker there might be the factory owner who argues that their demands are unreasonable, as well as other characters with differing perspectives such as political party activists, police officers, women's rights activists, and so on. The possibility to chat with different characters to explore multiple perspectives and reflect critically on their respective versions of the story could have obvious aesthetic and educational value. However, it would require careful design work in order to signal appropriately to the users that the characters they are interacting with might be unreliable. Such work might rely on literary scholarship about unreliable characters, but would be complicated by the great variability of LLM outputs - it would be hard to ensure that the chatbots adequately signal their own unreliability without breaking character.

If we can allow a museum chatbot to present unreliable characters and unreliable narration, we may also consider one final step on the "unreliability ladder": Can LLMs be used to design museum chatbots that are entirely unreliable, for the purpose of critique or satire? Satire can be used to expose hypocrisy and abusive power structures, to educate and to empower. An instructive example of educational satire in game form is the *Bad News Game* which was created as part of a research project to combat misinformation and "fake news" (Roozenbeek and van der Linden 2019). Paradoxically, the game invites players to take on the role of a producer of fake news, following the logic that learning about how fake news are created and disseminated might teach audiences how to better recognize and be critical of such disinformation techniques. This game predates the recent rise of LLMs, however it is easy to imagine how an LLM could be purposed for similar satirical uses. For instance, a science museum might invite visitors to chat with an LLM role playing as an oil industry lobbyist, in order to teach visitors about the rhetorical strategies used by the fossil fuel industry to counteract policy proposals to stop global warming. A historical museum might similarly invite visitors to chat with an LLM embodying historical values that are no longer acceptable in contemporary society, such as a slave-owner arguing for the merits of slavery, a 19th century man arguing for the subordination of women, or a 1970s white South African arguing for Apartheid. In principle such satirical LLMs might also be applied to the present, to facilitate satirical encounters with representatives of power in our conflict-ridden contemporary societies - if the politics of the museum insitution allows it to step into contemporary controversies.

## Concluding remarks

Critical play, like satire, relies on having a critical *edge* - a provocation of some sort, a message that aims to challenge preconceptions and inspire reflection; but which also may risk offending some people. Such provocations can be entertaining but also profoundly important for democratic discourse, by challenging power structures and allowing us to critically re-examine established beliefs. For the same reasons, designing for critical play may be risky for museums, which depend on public authorities as well as wealthy private funders.

Often, the funders of museum exhibitions are eager for museums to experiment with new technologies to facilitate engaging visitor experiences, in particular to reach young audiences. Calls for proposals for funding often reflect the latest technological hype, and in recent years AI and LLMs have been hyped to an unprecedented degree, with scenarios predicting extreme outcomes such as "intelligence revolution" (Cheok et al. 2024) or the end of humanity by the hands of a "superintelligence" (Bostrom 2014). However, at the time of writing this chapter (in May 2026) signs have started to appear of increasingly negative attitudes towards AI, with numerous news media reporting a public "backlash" against AI (Ramkumar et al. 2026; The British Broadcasting Corporation (BBC) 2026). Studies of audience's perception of AI-generated art have found that when people are told that something is created by AI they do tend to value it less than that which has been created by humans (Messer 2024; Raj et al. 2026; Neef et al. 2025). The use of generative AI for creative work has raised many concerns, with some going as far as to say that AI art is "theft" (Epstein et al. 2023; Goetze 2024). Such concerns are prevalent also in public discourse. Increasingly, text and images created by generative AI is derided as "slop" (Madsen and Puyt 2025; Shaib et al. 2025; “2025 Word of the Year” 2025; Silbey and Hartzog 2025). In this context, museums might be understandably skeptical about installing role playing chatbots deliberately designed to be unreliable narrators about historical events. The current controversies surrounding LLMs seem to draw viewpoints towards opposing poles. Designing LLM-based museum chatbots may require fending of both positive hyperbole as well as negative accusations in which all LLM outputs are seen as "slop" or "theft".

However, I believe it is worth the risk. Not just because of the great benefits of play, for museums and for its visitors; for learning as well as for its intrinsic value as a pleasurable activity that can engage children as well as adults. Critical play with LLMs may allow us to explore language, rhetoric, oral cultures and ideologies in interactive ways which go deeper than what can be gleaned by reading interpretive labels or listening to an audio guide. But perhaps even more importantly, by taking the weaknesses of LLMs seriously museums may offer a crucial opportunity for audiences to engage critically with LLMs as a technology. Given the ever-increasing impact AI technologies and technology corporations have on our society, it is extremely important to have venues which offer spaces both for playful exploration and critique. Such critical play may sometimes challenge our established notions of the past, as well as the authority of museums as knowledge institutions. But what is arguably just as important, in our contemporary society dominated by giant US technology corporations, is to facilitate playful critique of the technology and the values embedded in it. Teaching visitors to always be critical of the outputs from LLMs, and learning how to recognize "bullshit" even when wrapped in the seductive and versatile LLM style, is an immensely important and challenging task, as the numerous studies on overtrust cited in this chapter has demonstrated.

Sounds difficult? Don't worry, I asked the always supportive ChatGPT and it says you can do it, museums:

> *Perhaps the most exciting part is that museums do not need to have all the answers. The goal is not to present LLMs as trustworthy authorities, nor to dismiss them outright. It is to create spaces where people can explore these technologies together—curiously, critically, and sometimes playfully. Museums have always helped people navigate uncertainty. In that sense, working with LLMs is not a departure from their mission but a continuation of it.*[5]

You see? The AI says it, so it must be true.

# Author biography

Anders Sundnes Løvlie is Associate Professor and Head of the Media, Art, and Design research group at the IT University of Copenhagen. He has led an EU-funded research project on the design of "Hybrid Museum Experiences", and co-edited the book *Hybrid Museum Experiences: Theory and Design* with Annika Waern. He has published more than 80 scholarly publications focusing on experience design, artifical intelligence, human-computer interaction, museums, journalism and more.

[5] This paragraph was generated using chatgpt.com, prompted with the preceding paragraph as well as a request to formulate an ending to the article that was "encouraging and cheerful". This is the only piece of text in the book chapter that was generated with AI.